# Surface electronic structure and evidence of plain s-wave superconductivity in (Li$_{0.8}$Fe$_{0.2}$)OHFeSe

Y. J. Yan[1], W. H. Zhang[1], M. Q. Ren[1], X. Liu[1], X. F. Lu[3], N. Z. Wang[3], X. H. Niu[1], Q. Fan[1], J. Miao[1], R. Tao[1], B. P. Xie[1,2], X. H. Chen[3], T. Zhang[1,2*], D. L. Feng[1,2*]

[1] State Key Laboratory of Surface Physics, Department of Physics, and Advanced Materials Laboratory, Fudan University, Shanghai 200433, China

[2] Collaborative Innovation Center of Advanced Microstructures, Fudan University, Shanghai 200433, China

[3] Hefei National Laboratory for Physical Science at Microscale and Department of Physics, University of Science and Technology of China, Hefei, Anhui 230026, People's Republic of China

PACS number: 74.25.Ha, 74.25.Jb, 74.20.Rp, 68.37.Ef

*E-mail of T. Zhang: tzhang18@fudan.edu.cn
*E-mail of D. L. Feng: dlfeng@fudan.edu.cn

**(Li$_{0.8}$Fe$_{0.2}$)OHFeSe is a newly-discovered intercalated iron-selenide superconductor with a $T_c$ above 40 K, which is much higher than the $T_c$ of bulk FeSe (8 K). Here we report a systematic study of (Li$_{0.8}$Fe$_{0.2}$)OHFeSe by low temperature scanning tunneling microscopy (STM). We observed two kinds of surface terminations, namely FeSe and (Li$_{0.8}$Fe$_{0.2}$)OH surfaces. On the FeSe surface, the superconducting state is fully gapped with double coherence peaks, and a vortex core state with split peaks near $E_F$ is observed. Through quasi-particle interference (QPI) measurements, we clearly observed intra- and inter-pocket scatterings in between the electron pockets at the M point, as well as some evidence of scattering that connects Γ and M points. Upon applying magnetic field, the QPI intensity of all the scattering channels are found to behave similarly. Furthermore, we studied impurity effects on the superconductivity by investigating intentionally introduced impurities and intrinsic defects. We observed that magnetic impurities such as Cr adatoms can induce in-gap states and suppress superconductivity. However, nonmagnetic impurities such as Zn adatoms do not induce visible in-gap states. Meanwhile, we show that Zn adatoms can induce in-gap states in thick FeSe films, which is believed to have an $s_\pm$-wave pairing symmetry. Our experimental results suggest it is likely that (Li$_{0.8}$Fe$_{0.2}$)OHFeSe is a plain $s$-wave superconductor, whose order parameter has the same sign on all Fermi surface sections.**

# I. INTRODUCTION

The pairing mechanism is one of the pivotal issues in the study of iron-based superconductors [1,2]. Recently, heavily electron-doped iron selenide superconductors (HEDIS), such as $A_xFe_{2-y}Se_2$ (A=K, Rb, Cs…) [3-5] and single-layer FeSe films on $SrTiO_3$ (STO) [6-11], have attracted tremendous interest. In these materials, the absence of hole-like Fermi surfaces, together with the nodeless superconducting gap [5-9], greatly challenges existing theories, especially the weak coupling theories that were rather successful in predicting the $s_\pm$ pairing symmetry in iron pnictide superconductors [12,13]. Currently, there is no consensus regarding the pairing symmetry of HEDIS, and many forms of pairing symmetry have been proposed, including $s_{++}$-wave, *d*-wave, bonding-antibonding $s_\pm$-wave *etc.* [13-18]. In the case of single layer FeSe/STO films, our recent STM study showed that system to be a plain *s*-wave superconductor [19]. However, this may be a special case, since signatures of strong interfacial electron-phonon interactions have been observed, which may play a dominant role in this interfacial superconducting system [10].

Recently, an intercalated FeSe-derived superconductor, $(Li_{0.8}Fe_{0.2})OHFeSe$, has been synthesized which exhibits superconductivity above 40 K [20]. This material consists of alternating FeSe and $(Li_{0.8}Fe_{0.2})OH$ layers. Preliminary ARPES measurements indicated that it is also heavily electron doped with Fermi surfaces only at the M points, and there are no side bands induced by strong electron-phonon interactions as observed in single layer FeSe/STO [21, 22]. Because this material does not have intrinsic phase separation and is air stable, it is a promising candidate to study the superconductivity of HEDIS.

STM is a powerful tool for studying superconductivity and gaining phase information of the order parameter [23-28]. In this paper, we report a systematic STM study on $(Li_{0.8}Fe_{0.2})OHFeSe$ single crystals. We observed two kinds of surface terminations on the cleaved sample, which we identified to be the FeSe and $(Li_{0.8}Fe_{0.2})OH$ surfaces. On the FeSe surface, we observed fully gapped tunneling spectra with double coherence peaks, which is similar to those of single-layer FeSe/STO [6]. Magnetic vortices were found to be spatially isotropic with double-peaked bound states at the core. On the $(Li_{0.8}Fe_{0.2})OH$ surface, a metallic state is observed without obvious superconducting gap opening. QPI patterns on the FeSe surface revealed intra- and inter-pocket scattering of the electron pockets at M points. Meanwhile, a feature centered at $(\pi, 0)$ which may correspond to Γ-M scattering is observed near Fermi energy. To explore the pairing symmetry, we measured the magnetic field dependence of the QPI, as well as the impurity effects on intentionally introduced impurities (deposited Cr and Zn atoms) and intrinsic defects. We found that, 1: all the scattering channels in QPI behave similarly under magnetic field, including the feature at $(\pi, 0)$; and 2: magnetic impurities such as Cr adatoms can induce in-gap states and locally suppress the superconductivity on FeSe surface, while the non-magnetic Zn adatoms do not induce any visible in-gap states. Furthermore, we checked the effect of Zn atoms deposited on thick FeSe films (which is believed to have an $s_\pm$-wave symmetry [29, 30]) and found that they *do* induce pronounced in-gap states in this case. We show that although a thorough understanding of the QPI and impurity effects will need more theoretical works with considering the microscopic details, our results do not show evidence of sign change in $(Li_{0.8}Fe_{0.2})OHFeSe$. It is likely that $(Li_{0.8}Fe_{0.2})OHFeSe$ is a plain *s*-wave superconductor whose order parameter has the same sign on all Fermi surface sections.

## II. METHODS

$(Li_{0.8}Fe_{0.2})OHFeSe$ single crystals were grown by a novel hydrothermal method as described in ref. 20. Temperature dependence of the electrical resistivity under zero magnetic field was measured by a standard DC four-probe method using Quantum Design PPMS. Temperature dependence of the magnetic susceptibility was measured using Quantum Design MPMS. The STM experiment was conducted in a cryogenic STM with a base temperature of 0.4 K. The sample was cleaved in vacuum at 77 K and immediately transferred into the STM module. FeSe films (25 monolayers thick) were grown by co-deposition of high purity Se (99.999%) and Fe (99.995%) on graphitized SiC (0001) held at 620 K. The graphitized SiC (0001) substrates were prepared by direct heating of SiC (0001) at 1650 K. STM measurements were taken at 4.2 K for $(Li_{0.8}Fe_{0.2})OHFeSe$ and at 0.4 K for FeSe films. Cr and Zn atoms were evaporated onto the surface at low temperatures (~50 K). PtIr STM tips were used after being treated on Au or Ag surface. dI/dV spectroscopy was collected using a standard lock-in technique with modulation frequency f = 973 Hz and typical modulation amplitude (ΔV) 1 mV.

## III. RESULTS AND DISCUSSION

### A. Resistivity and magnetic susceptibility

Temperature dependence of the resistivity and DC magnetic susceptibility of $(Li_{0.8}Fe_{0.2})OHFeSe$ single crystals have been measured, as shown in Fig. 1. A sharp superconducting transition at about 40 K is observed both in Figs. 1(b) and 1(c), confirming the good quality of the $(Li_{0.8}Fe_{0.2})OHFeSe$ crystal.

### B. Surface topography and tunneling spectrum

As illustrated in Fig. 1(a), $(Li_{0.8}Fe_{0.2})OHFeSe$ adopts a structure with alternate stacking of anti-PbO-type FeSe and $(Li_{0.8}Fe_{0.2})OH$ layers, with an in-plane lattice constant of 3.78 Å [20]. The natural cleavage would expose either FeSe or $(Li_{0.8}Fe_{0.2})OH$ terminated surfaces. In our STM study, indeed two kinds of surface terminations have been observed, as shown in Figs. 2(a) and 2(b). Judging from their topographic and spectroscopic characters (as shown throughout this paper), we attribute Fig. 2(a) as the FeSe-terminated surface and Fig. 2(b) as the $(Li_{0.8}Fe_{0.2})OH$-terminated surface. The FeSe surface is atomically flat with two kinds of intrinsic defects, as marked by I and II. Enlarged images of these are shown in Figs. 2(c) and 2(d), respectively. Type II defects are located at Se sites, and are likely to be Se vacancies, as reported for thick FeSe films [29]. Type I defects are dimer-like, with the center located at the Fe site. They could be Fe vacancies [31] or substitutional impurities at the Fe site [32]. The (Li$_{0.8}$Fe$_{0.2}$)OH surface is rougher than the FeSe surface, as shown in Fig. 2(b), probably due to the high level of Fe substitution for Li. Nevertheless, the atomic lattice can still be resolved (Fig. 2(b) inset), with a lattice constant the same as that of the FeSe surface. In addition, tunneling barrier heights were mapped to identify (Li$_{0.8}$Fe$_{0.2}$)OH- and FeSe- terminated surfaces. The tunneling current $I$ is expected to decay exponentially with the tip-sample distance $z$ as $I \propto exp(-Z\sqrt{8m\Phi/\hbar^2})$, where $\Phi$ is the local barrier height (LBH) that gives an estimation of work function. Averaged $I(z)$ curves measured on both surfaces with the same tip are shown in Fig. 2(e), which yields LBHs of 2.7 eV for the (Li$_{0.8}$Fe$_{0.2}$)OH surface and 3.42 eV for the FeSe surface. This provides another way to distinguish these two surfaces.

Fig. 2(f) displays typical dI/dV spectra taken on the two surface terminations. For the FeSe surface (red curve), a fully-developed superconducting gap with double coherence peaks at ±9 mV and ±15 mV is observed. The gap bottom with nearly zero tunneling conduction is 5 meV wide. It is remarkable to note that both the gap structure and gap size are similar to that observed in single-layer FeSe/STO [6]. For the (Li$_{0.8}$Fe$_{0.2}$)OH surface (blue curve), the spectrum shows metallic behavior with a weak dip at Fermi level. This indicates that the (Li$_{0.8}$Fe$_{0.2}$)OH surface is metallic but likely *not* superconducting, and implies that the coupling between FeSe and (Li$_{0.8}$Fe$_{0.2}$)OH layers is rather weak.

### C. Magnetic vortex

The superconductivity in the FeSe surface is further investigated by imaging magnetic vortices. Fig. 3(a) shows a zero bias conductance (ZBC) mapping of a 50×50 nm$^2$ area, measured at B = 11 T. The vortices are clearly visible; however, the vortex lattice is highly disordered. By comparing with topography (see Fig. s1(a)), we found that this is because the dimer-like defects are strong pinning centers (they locally suppress superconductivity, as shown later). The pinned vortices (shown by yellow circles) have a different appearance than "free" vortices in the ZBC mapping, as highlighted in Fig. 3(a). For this reason, we only studied free vortices as marked by white dashed circles. The overall shape of a single free vortex is spatially isotropic (see Fig. 3(b) inset), which differs from the elongated vortices in FeSe thick films [29]. An exponential fit to the line profile of the Fig. 3(b) inset gives an estimate for the Ginzburg-Landau coherence length ($\xi$) of 2.3 nm.

Fig. 3(c) shows the spectrum taken at the center of a free vortex core (red curve) – a pair of peaks at energies of ±2.1 meV with asymmetric intensities is observed. Away from the core center, these two peaks shift to higher energy and eventually merge into the gap edge (Fig. 3(d), shown in false color), and the double-gap structure is recovered (blue curve in Fig. 3(c)). The presence of core state with a pair of peaks is consistent with fully gapped superconductivity, and indicates that the system is in the "quantum limit" [33-35], where the thermal smearing is sufficiently low that one can resolve the bound states having energies $E_p = \pm \Delta^2/2E_b$ ($E_b$ is the occupied band width, see more discussion in Supplementary Information part I). By using the larger gap value of $\Delta = 15$ meV and $E_p = 2.1$ meV, we get $E_b = 53$ meV. This agrees well with the occupied width of the electron bands at M measured by ARPES [21, 22] and our QPI data shown below.

### D. Electronic structure

To further examine the electronic structure of (Li$_{0.8}$Fe$_{0.2}$)OHFeSe, we performed dI/dV mappings to reveal the QPI patterns. A set of dI/dV maps within the energy range of ±30 meV were carried out in a 35×35 nm$^2$ area on FeSe surface, several of them are shown in Figs. 4(a-h) (Mapping area is the same as that shown in Fig. 2(a), see Fig. s2 for complete set of dI/dV maps), which clearly show interference modulation around defects. Fig. 4(i) shows the comparison of the QPI intensities at $V_b$ = 6 meV and -6 meV along the same linecut crossing the same defect. The anti-phase relation of the QPI modulation near the defect can be seen, which is a characteristic of Bogoliubov quasiparticles.

Figs. 5(a-h) show the fast Fourier transforms (FFT) of the dI/dV maps in Figs. 4(a-h). The FFTs are fourfold symmetrized to reduce noise since all the QPI patterns were found to be fourfold symmetric (see also Fig. s2 for the raw FFTs). The common features throughout Figs. 5(a-h) are ring-like patterns centered at (0, 0), (π, π) and (0, 2π) (marked as Ring 1, Ring 2, and Ring 3, respectively), which have also been observed in single-layer FeSe/STO [19, 36]. These features originate from intra- and inter-pocket scattering of the electron pockets at the M points of the Brillouin Zone (BZ), as sketched in Fig. 5(i). In Fig. 5(j), we show the simulated FFT from calculating the joint density of states (JDOS), based on the *unfolded* Fermi surface shown in Fig. 5(i) (solid curves). One found that the ring-like features are well reproduced in the simulation; while the fourfold anisotropy of Ring 1 and the oval shape of Ring 3 (as clearly seen in Figs. 5(b-d)) can also be reproduced by considering finite ellipticity of the electron pockets. Fitting to Ring 3 in Fig. 5(b) yields an ellipse with a long to short axis ratio of 1.1. We found that the unfolded Fermi surface can actually produce better FFT simulation than the folded Fermi surface, which likely means the BZ folding effect is *weak* in this system (see Fig. s4 for more details). According to the simulation, the radius of Ring 2 is twice of the averaged radius of the electron pockets. In Fig. 5(k), we show the azimuth-averaged FFT line cuts surrounding (π, π) (as marked in Fig. 5(b)), taken at various energies outside the gapped region. The dispersion of Ring 2 is clearly seen and parabolic fitting yields a band bottom at -50 (±5) meV and an averaged Fermi crossing ($k_F$) of 0.21 (±0.1) Å$^{-1}$. These agree well with the ARPES measurements [21, 22] and are close to the values of single-layer FeSe/STO [7-9] (band bottom at -60 meV and $k_F$ = 0.22 Å$^{-1}$). One may notice that Ring 2 appears to split into two rings at E < -10 meV (see Figs. 5(h), 5(k) and Fig. s2). The inner ring follows the parabolic dispersion but the outer ring is almost non-dispersive. The origin of this splitting is unclear and needs further investigation.

Interestingly, besides the ring-like structures, there are other features in the QPI on the FeSe surface. From Figs. 5(c-h) and the FFT line cuts summarized in Fig. 5(l), one sees that below E = 20 meV, some feature shows up around ($\pi$, 0) with increasing intensity as the energy decreases. It is clearly separated from other scattering channels and persists across $E_F$. As sketched in Fig. 5(i), the position of this scattering corresponds to $q_4$, which connects the $\Gamma$ and M points. Since this feature disappears above 20 meV and becomes pronounced at low energies, it likely arises from scattering between the hole-like pocket at $\Gamma$ and the electron pocket at M. However, ARPES data on (Li$_{0.8}$Fe$_{0.2}$)OHFeSe indicates that the top of the hole band at $\Gamma$ is 50 meV below $E_F$, and there are no other bands crossing $E_F$ around $\Gamma$ [21,22]. Thus the origin of this possible $\Gamma$-M scattering is puzzling. One may speculate the hole band at $\Gamma$ may still have some residual weight near $E_F$, which could be due to the broadening of impurity scattering. We noticed that some theoretical works suggest such band without clear Fermi level crossing ("incipient band") may still play important role on superconductivity [37, 38].

QPI measurements on (Li$_{0.8}$Fe$_{0.2}$)OH surfaces were also performed. Figs. 6(a-d) are selected dI/dV maps in a 120×120 nm$^2$ area, the corresponding FFT images are shown in Figs. 6(e-h) (see Fig. s3 for a complete set of dI/dV maps and FFTs). A single, circular scattering ring is observed around (0, 0), without any other high-$q$ features. Meanwhile, the size of the ring decreases with decreasing energy, indicating that it is likely from the intra-band scattering of a 2D electron pocket. In Fig. 6(g), we summarize the FFT line cuts though the center of the scattering ring. A parabolic fit to the dispersion yields a band bottom at -50 meV and a Fermi crossing at 0.09 Å$^{-1}$. Since the (Li$_{0.8}$Fe$_{0.2}$)OH layers act as an electron reservoir that provides electrons to the FeSe layers, the existence of an electron pocket on this surface could be expected. We note that such a Fermi pocket is not observed in the ARPES studies, which may be due to the negligible photoemission matrix element of these states [21, 22].

So far our measurement confirms well-developed superconductivity at the FeSe surface. It is expected that the FeSe layer may lose half of its bulk electron carriers after cleavage, however the similar observed band structure to that of single-layer FeSe/STO and the well-developed superconducting gap indicate that the exposed FeSe layer is still sufficiently doped. The absence of a superconducting gap on the (Li$_{0.8}$Fe$_{0.2}$)OH surface indicates weak coupling between the (Li$_{0.8}$Fe$_{0.2}$)OH and neighboring FeSe layers. Thus the FeSe-terminated surface structurally resembles single-layer FeSe/STO, except that the STO substrate is now replaced by the (Li$_{0.8}$Fe$_{0.2}$)OH layer, which decouples the top FeSe layer from the bulk. The feature observed here apart from single-layer FeSe/STO is the possible $\Gamma$-M scattering at ($\pi$, 0). As theoretically predicted, such a scattering channel may lead to a sign-changing $s_\pm$-wave pairing (even the $\Gamma$ band doesn't have a Fermi surface) [12, 13, 38]. Thus it would be important to check the pairing symmetry of this system.

### E. Magnetic field dependent QPI

In STM study, one way to gain phase information of superconducting order parameter ($\Delta_k$) is to check the magnetic field dependence of the QPI. In previous study of cuprate [25], it is found that in presence of magnetic vortices, the scatterings which preserve the sign of $\Delta_k$ were enhanced, and the scatterings which change the sign of $\Delta_k$ were suppressed. Similar effect has also been observed in $FeTe_xSe_{1-x}$, which is believed to have an $s_\pm$-wave pairing [30]. Further theoretical works show that the disordered vortex cores which locally suppressed the order parameter [39, 41], and/or the impurities insides of the vortex core which acquire additional resonant or Andreev scattering [40], can indeed enhance the sign-preserving scattering channel. Meanwhile the strength of sign-changing scattering is likely not directly affected by vortices; A weak, overall suppression of all the scatterings may be expected, due to the additional phases acquired by quasi-particles moving through disordered vortex lattice [31]. Therefore, one may still expect that the sign-changing and sign-preserving scatterings will show different intensity change under magnetic field. In our case, the observed vortex lattice (Fig. 3(a)) is significantly disordered and a part of the vortices are pinned around defects, which satisfies the condition discussed in refs. 39-41.

We then carried out dI/dV mapping under magnetic fields of 0 T and 11 T in the same scan area (32×32 nm$^2$) within the energy range of ±30 meV. Figs. 7(a-d) show dI/dV maps and their FFTs taken at $V_b$ = 12 meV under B = 0 T and 11 T for comparison (see Fig. s5 for comparisons at more energies). In Fig. 7(e), we show the difference of the QPI intensities between Figs. 7(c) and 7(d). Here we intentionally suppressed the intensity near (0, 0) and all the Bragg spots, because they are either irrelevant to QPI or could introduce artifacts. One sees that apparently an overall suppression occurs for all scattering channels. We then compare the relative change of the intensities of different scattering channels, as a function of energy. The scattering intensities of each channel are obtained through integrating relevant areas in the FFT maps (shaded areas in Fig. 7(c)), again excluding the regions near (0, 0) and the Bragg spots. As shown in Fig. 7(f), all the scattering channels show similar suppression in the amplitude when the energy approaches the gap edge, including the possible Γ-M scattering ($q_4$). Thus, despite the overall suppression requires further quantitative explanation, no evidence is observed as an indication of sign-changing scatterings here.

### F. Impurity effect

Besides the QPI measurement, impurity-induced effects are another way to explore the pairing symmetry. In general, the response of superconductivity to local impurities depends on the pairing symmetry and the characteristic of the impurities [26]. It is known that for *s*-wave pairing, only magnetic impurities can break the Cooper pair and induce in-gap bound states [42]. However, for phase-changing pairing symmetries such as *d*-wave and $s_\pm$-wave, it is predicted that non-magnetic impurities with proper scattering potentials, can also induce in-gap states and suppress superconductivity [43-45], which is supported by STM measurements on cuprates [28], NaFeAs [32] and LiFeAs [46]. Meanwhile, several theoretical works have shown that non-magnetic impurities can also help to identify the pairing symmetry of

$K_xFe_{2-y}Se_2$ [47-49], which has similar band structure with $(Li_{0.8}Fe_{0.2})OHFeSe$.

We investigated the impurity effect in $(Li_{0.8}Fe_{0.2})OHFeSe$, by controllably introducing impurities on the FeSe surface, as well as by studying the intrinsic defects. In the first case, impurity atoms Cr (magnetic) and Zn (non-magnetic) were deposited separately onto the sample holding at low temperature (~50 K). These atoms appear as bright protrusions on the FeSe surface in the topography (Figs. 8(a-b)). Assuming the interaction between the low-temperature adsorbed atoms and underlying FeSe lattice to be weak, the impurity atoms are expected to retain their magnetic/non-magnetic character after adsorption. In Figs. 8(c-d), we show local tunneling spectra near Cr and Zn atoms. On the Cr site, the superconducting gap is greatly suppressed and a pair of asymmetric peaks appear in the gap. These are hallmarks of impurity-induced in-gap states. Away from the Cr site, the impurity states are weakened and the superconducting gap gradually recovers. Meanwhile, for a Zn impurity, the superconducting gap size remains unchanged at and near the Zn site. Although the coherence peaks at ±9 meV change in intensity near Zn sites, there is no evidence of in-gap states.

The absence of in-gap states at non-magnetic impurities intuitively suggests an s-wave pairing without sign change. However, because Fe-based superconductors are multiband systems, recent theoretical works show that the formation of sharp in-gap states on non-magnetic impurities is not only subject to pairing symmetry, but also highly depends on the details of band structure and the strength of scattering potentials [50]. Here we are not going to give a theoretical calculation considering all the details of $(Li_{0.8}Fe_{0.2})OHFeSe$ band structure and the scattering potentials of Zn adatoms, which could be difficult to determine. Instead we performed a comparison experiment - checking the impurity effect of Zn adatom on undoped FeSe, which is widely believed to have an $s_\pm$-wave pairing [29]. We grew 25 ML thick FeSe film on SiC substrate (see Methods section), and Zn atoms were deposited on such film the same way as for $(Li_{0.8}Fe_{0.2})OHFeSe$. Figs. 8(e) and 8(f) show the topography and local tunneling spectra near a Zn adatom on thick FeSe films, respectively. One can clearly see that at the Zn site the superconducting gap of undoped FeSe film is dramatically suppressed and a pair of in-gap states emerge at ±1.2 meV. The presence of in-gap states strongly supports the phase-changing pairing in undoped FeSe, and indicates Zn adatoms are effective scatterers for this multiband system. The remarkable different response of $(Li_{0.8}Fe_{0.2})OHFeSe$ to the same Zn impurity could arise from the pairing symmetry as well as it's different electron structure to undoped FeSe.

We also measured the impurity effects induced by intrinsic defects (Fig. 9). Fig. 9(c) shows the tunneling spectra taken near a dimer-like type I defect in Fig. 9(a). At the center of the dimer, in-gap states with asymmetric intensities at ±3 meV are observed, and the superconducting gap is almost completely suppressed. Since the type I defects should be Fe vacancies or substitutional impurities on the Fe site, they are likely to carry spin and be magnetic. We note that in $K_xFe_{2-y}Se_2$, Fe vacancies have been experimentally proven to be magnetic and induce in-gap states [31]. The strong local suppression of superconductivity makes type I defects effective pinning sites for magnetic vortices. Tunneling spectra for type II defects, which we attribute to Se vacancies, are shown in Fig. 9(d). In contrast to type I defects, the superconducting gap is unaffected at the defect (Se) site and nearby, and no in-gap states are observed. We noticed that Se vacancies in thick FeSe films *do* induce in-gap states and suppress superconductivity, as reported previously in ref. 29. Thus it gives another instance that the same type of impurity can have different effects in undoped FeSe and $(Li_{0.8}Fe_{0.2})OHFeSe$. One may speculate the Se vacancies are non-magnetic, which play a similar role as Zn adatoms. Overall, the impurity effects we observed are consistent in that only magnetic impurities induce in-gap states in $(Li_{0.8}Fe_{0.2})OHFeSe$, while non-magnetic ones do not.

Upon finishing this manuscript, we noticed a theoretical work shows that non-magnetic impurities may not induce observable in-gap states for "incipient" $s_{\pm}$-wave pairing [51]. In this scenario the gap changes sign on the incipient hole band which does *not* have a Fermi surface. Identification of such sign change may require more phase sensitive methods beyond the impurity effect.

## IV. CONCLUSION

Overall, our STM study revealed distinct electron structure on FeSe and $(Li_{0.8}Fe_{0.2})OH$ terminated surface of $(Li_{0.8}Fe_{0.2})OHFeSe$, and offers several independent hints for the pairing symmetry. Fully-gapped tunneling spectra and double-peaked vortex core states indicate a nodeless superconducting state. Magnetic field dependence of QPI and impurity effects did not show sign of phase change, although some evidence of Γ-M scattering is observed. Magnetic impurities can induce in-gap states but nonmagnetic ones such as Zn adatoms do not. These results together would suggest a plain *s*-wave pairing in $(Li_{0.8}Fe_{0.2})OHFeSe$, whose order parameter has the same sign on all Fermi surface sections. Previously we have shown single-layer FeSe/STO is also a plain *s*-wave superconductor, where side bands due to strong interfacial electron-phonon interactions was observed [10]; while the side bands are absent in the ARPES data of $(Li_{0.8}Fe_{0.2})OHFeSe$ [22]. All these findings show that the $s_{++}$ pairing symmetry is likely a robust feature of HEDIS. This finding is consistent with strong-coupling theories based on local antiferromagnetic coupling [14, 17], or orbital-fluctuation-mediated pairing mechanism [52].

*Note added*: upon finishing this work, we noticed another independent STM study on $(Li_{0.8}Fe_{0.2})OHFeSe$ in the ref. 53.

## ACKNOWLEDGEMENTS

We thank Professors J. P. Hu, Prof. Y. Chen and Dr. Y. W. Guo for helpful discussions. This work was supported by the National Science Foundation of China, and National Basic Research Program of China (973 Program) under the grant No. 2012CB921402, No. 2011CBA00112, and No. 2011CB921802.

## Figures for the main text:

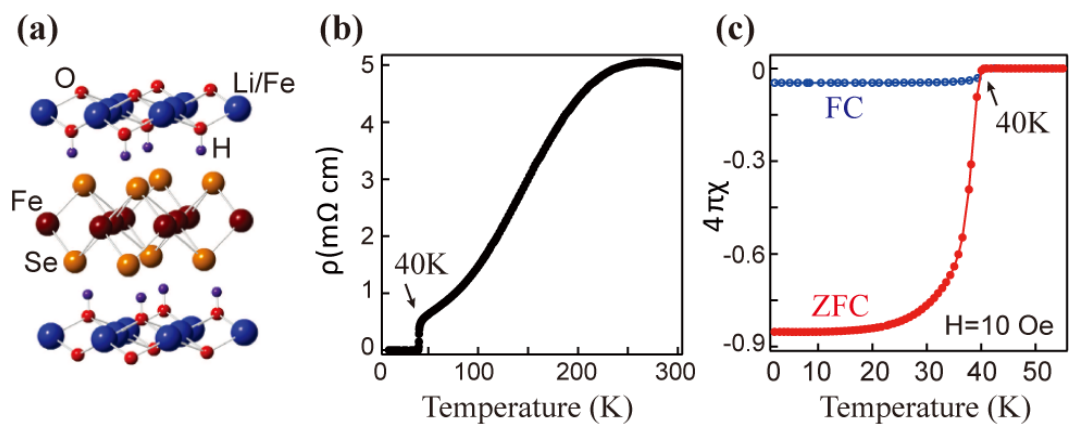

FIG. 1. (**a**) Schematic of the crystal structure of $(Li_{0.8}Fe_{0.2})OHFeSe$, (**b**) Temperature dependence of the resistivity of $(Li_{0.8}Fe_{0.2})OHFeSe$ single crystal. (**c**) Temperature dependence of the DC magnetic susceptibility of $(Li_{0.8}Fe_{0.2})OHFeSe$ measured through zero-field cooling (ZFC) and field cooling (FC).

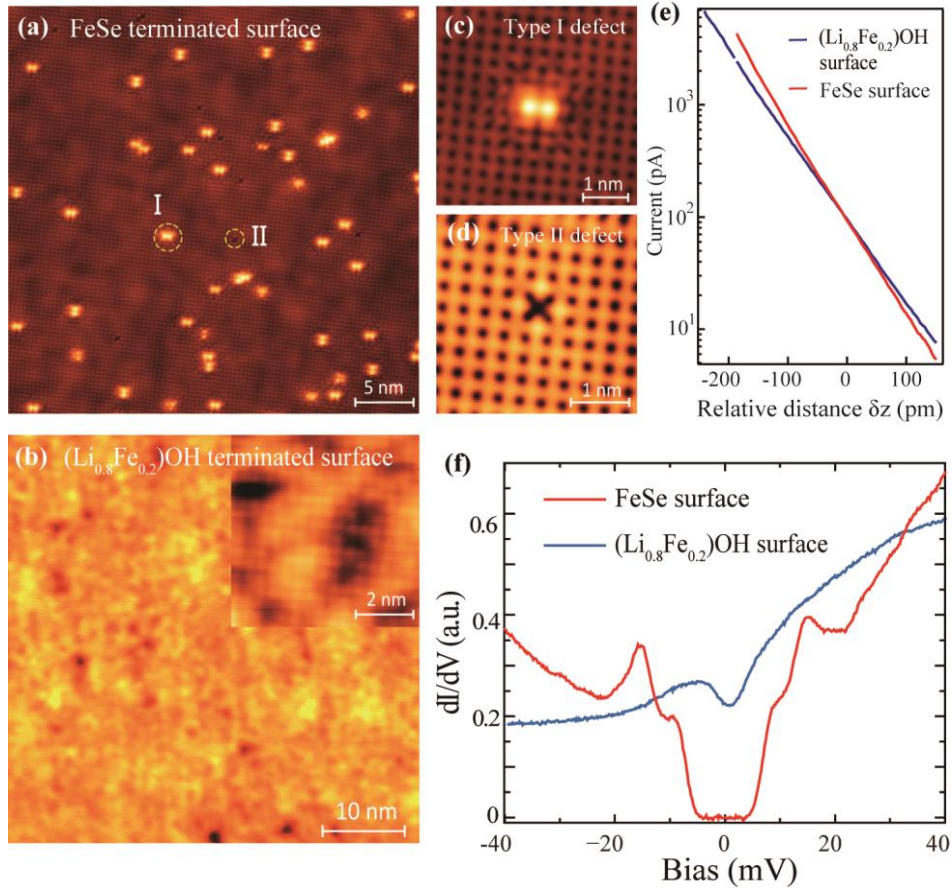

FIG. 2. Surface topography and dI/dV spectra of $(Li_{0.8}Fe_{0.2})OHFeSe$ single crystal. (**a**) Typical topographic image on the FeSe-terminated surface (bias voltage: $V_b$= 50 mV, current: $I$= 10 pA). Two types of defects (I, II) are marked, and expanded views of their morphologies are shown in (**c**) and (**d**). (**b**) Typical topographic image on the $(Li_{0.8}Fe_{0.2})OH$-terminated surface ($V_b$ =100 mV, $I$ = 50 pA). Inset is the atomically resolved image ($V_b$ = 5 mV, $I$ = 300 pA). (**e**) $I(z)$ curves measured on $(Li_{0.8}Fe_{0.2})OH$- and FeSe-terminated surfaces ($V_b$ = 200 mV, $I$ = 100 pA). (**f**) Averaged superconducting gap spectra on FeSe- and $(Li_{0.8}Fe_{0.2})OH$-terminated surfaces ($V_b$ =40 mV, $I$ = 150 pA, $\Delta V$= 1 mV).

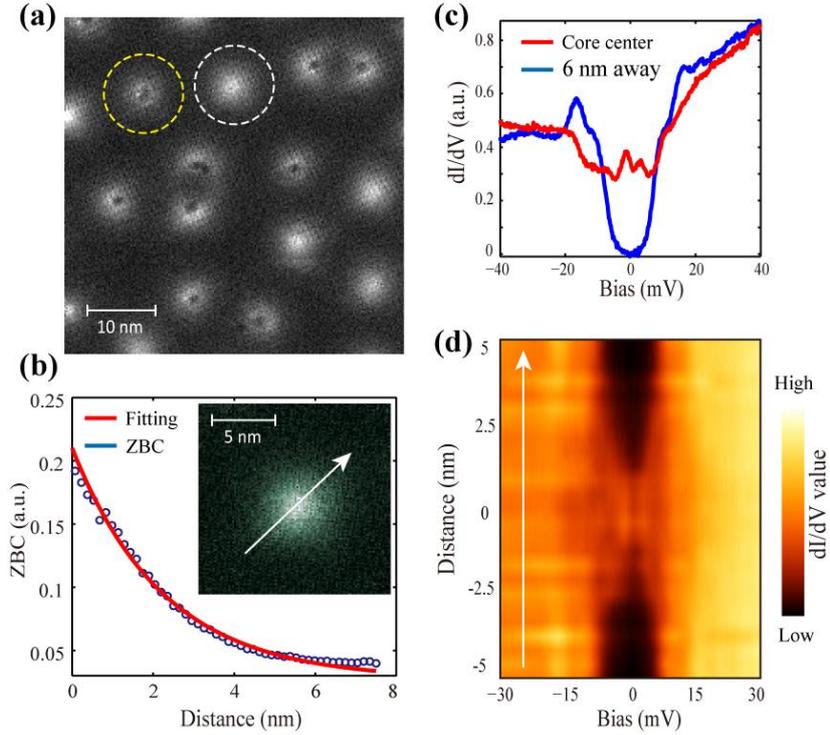

FIG. 3. Vortex mapping on FeSe-terminated surface. (**a**) Vortex mapping on FeSe-terminated surface at $V_b$=0 mV under 11 T magnetic field. The vortex in the yellow circle with suppressed intensity near the core center are pinned by the dimer-like defects, while the vortex in the white circle is a "free" vortex which is not pinned. (**b**) Exponential fit to the line profile of a single free vortex in the ZBC mapping. Inset: a zoomed-in ZBC map of a single free vortex. (**c**) dI/dV spectra taken at the vortex core center and 6 nm away from the center ($V_b$ =40 mV, $I$ = 100 pA, $\Delta V$= 1 mV). (**d**) Evolution of the dI/dV spectra taken along the line across the free vortex core, as marked in the inset of (**b**), shown in false color.

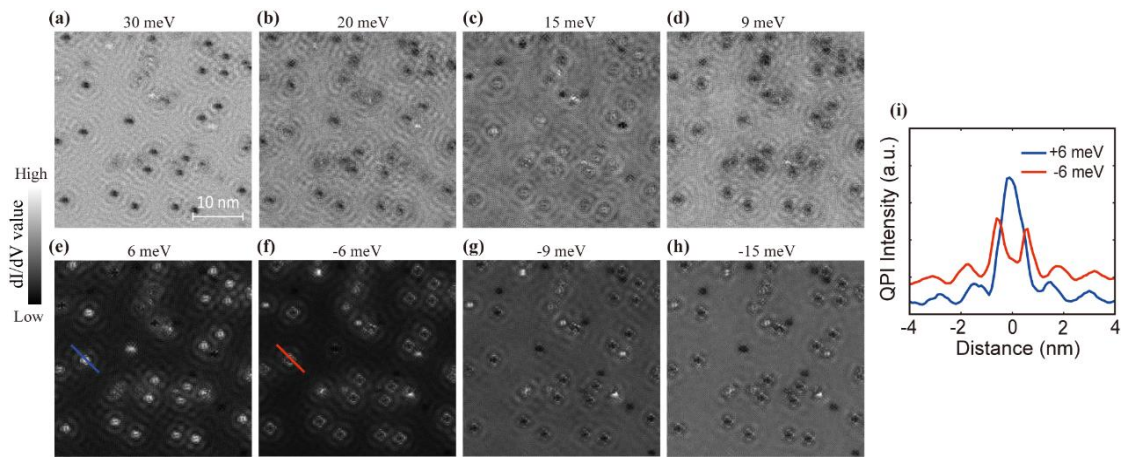

FIG. 4. (**a-h**) dI/dV maps on FeSe-terminated surface taken at different bias voltage, in the same 35×35 nm² area. Set point: $V_b$ = 30 mV, $I$ = 150 pA. (**i**) Linecut profiles taken along the line shown in (**e**) and (**f**)

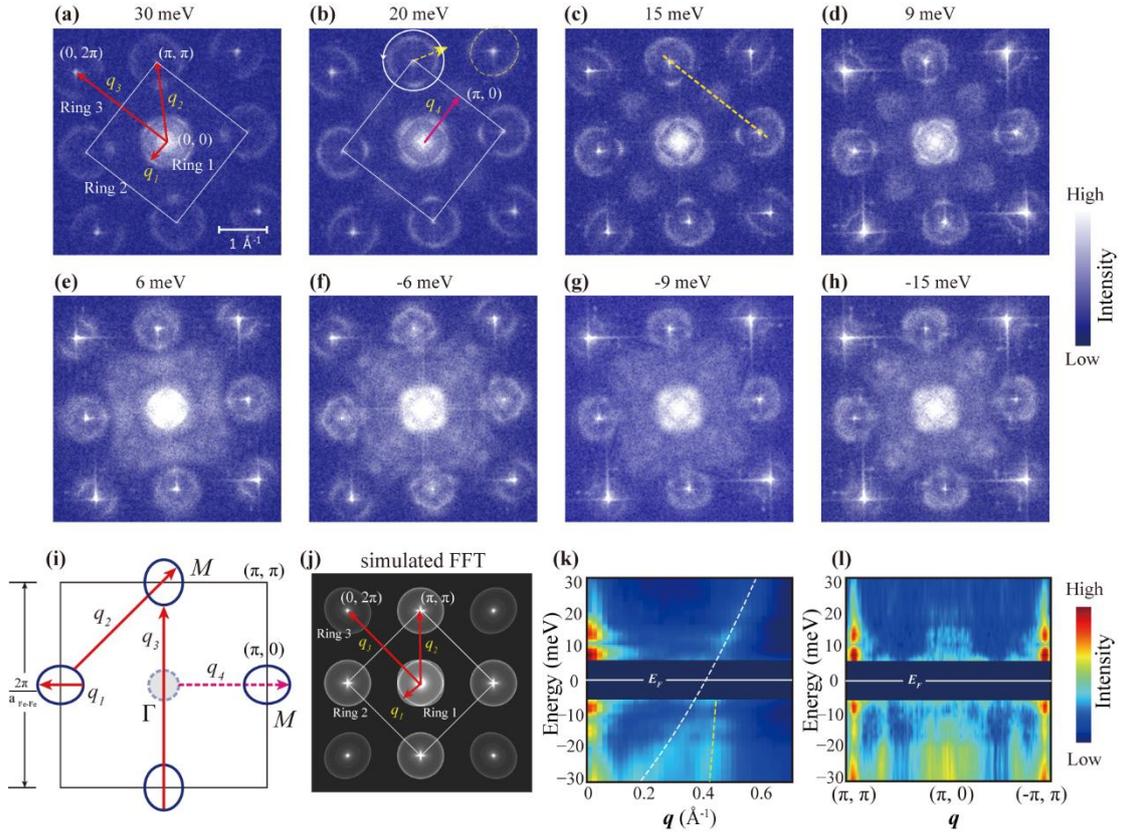

FIG. 5. QPI patterns on FeSe- terminated surfaces. (**a-h**) FFT transformations of the dI/dV maps shown in Fig. 4(a-h). The white square represents the unfolded Brillouin zone. Different scattering channels are indicated by the red arrows. (**i**) Schematic of the unfolded Brillouin zone and Fermi surface of the FeSe surface. The dashed circle indicates the possible residual weight of hole-like pocket at Γ. The possible scattering channels are marked by $q_1 \sim q_4$. (**j**) Simulated FFT corresponding to the Fermi surface shown in (**i**) (without the pocket at the Γ point). (**k**) FFT line cuts extracted from the yellow dashed arrow in (**b**) and azimuth-averaged with respect to (π, π), taken at various $V_b$ and shown in false color. The white dashed curve is a parabolic fit to the dispersion of Ring 2. The yellow dashed curve indicates the splitting of Ring 2 below -10 meV. (**l**) The FFT line cuts extracted along the yellow dashed line in panel (**c**), taken at various $V_b$ and shown in false color.

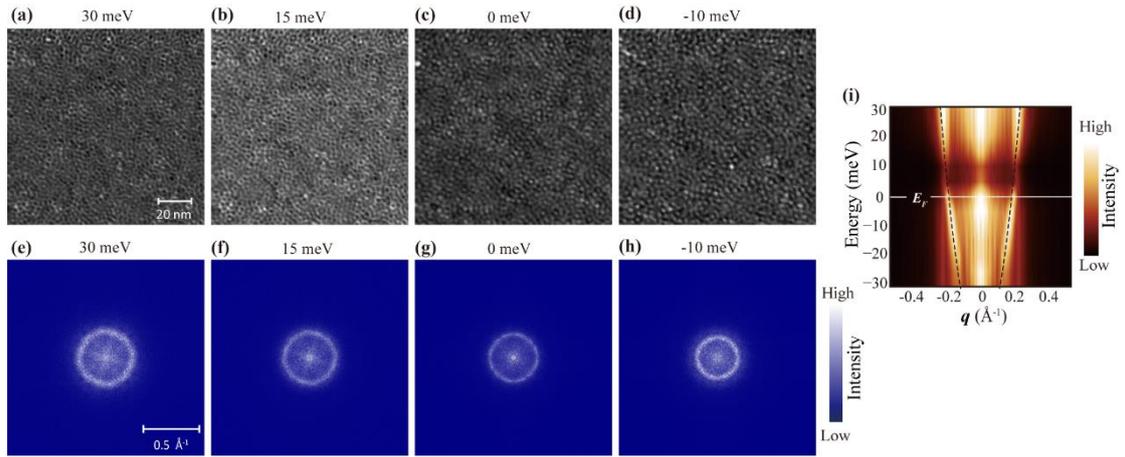

FIG. 6. QPI patterns on $(Li_{0.8}Fe_{0.2})OH$- terminated surfaces. (**a-d**) dI/dV maps taken in a 120×120 $nm^2$ area of the $(Li_{0.8}Fe_{0.2})OH$-terminated surface. The mapping energies are labeled in the images. Each map has 400×400 pixels. (**e-h**) FFTs of the dI/dV maps shown in panels (**a-d**) (four-fold symmetrized). (**i**) Line cuts extracted from the FFTs of the $(Li_{0.8}Fe_{0.2})OH$ surface at various $V_b$, shown in false color. The black dashed curve is a parabolic fit to the dispersion of the scattering ring, which indicates an electron pocket.

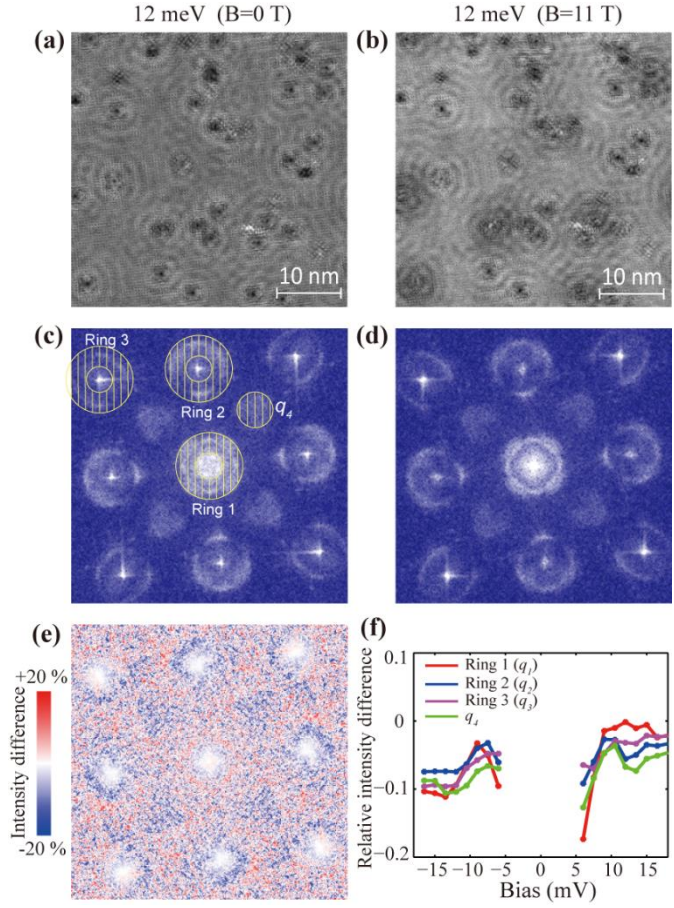

FIG. 7. Magnetic field dependence of QPI patterns on FeSe-terminated surface. (**a-b**) dI/dV maps taken under magnetic fields of 0 T and 11 T in the same scan area at $V_b$ =12 mV. Set point: $V_b$ = 30 mV, $I$ = 150 pA. (**c-d**) FFTs of the dI/dV maps shown in (**a-b**), respectively. Masked areas in (**c**) show the integration windows for different scattering channels. (**e**) The difference in QPI intensities at $V_b$ =12 mV, which is calculated by $(FFT_{11T} - FFT_{0T}) / (FFT_{11T} + FFT_{0T})$. $FFT_{11T}$ and $FFT_{0T}$ are shown in (**d**) and (**c**), respectively. The intensity near (0, 0) and all Bragg spots is suppressed by a factor of $1-\Sigma[Gaussian(q_{(Bragg)}, \sigma)]$. All scattering channels are suppressed under high magnetic field. (**f**) The relative change of the intensities of different scattering channels, as a function of energy. The scattering intensities of each channel were obtained through integrating the relevant area in the FFT maps, as shown in (**c**).

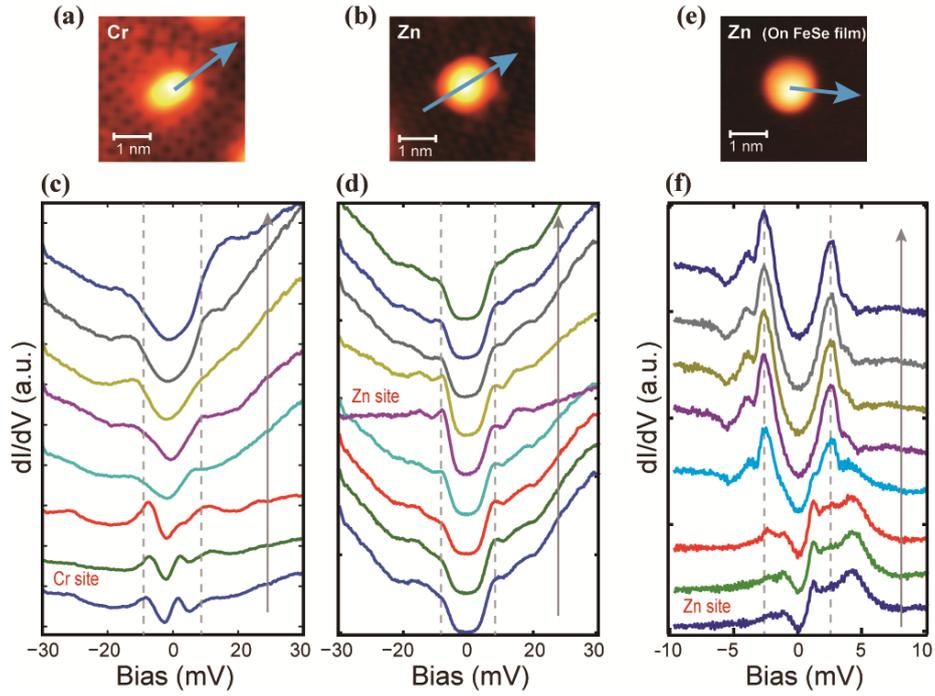

FIG. 8. Impurity-induced effects on the superconductivity of (Li$_{0.8}$Fe$_{0.2}$)OHFeSe and thick FeSe films. (**a-b**) Topographic images of single Cr and Zn adatoms on FeSe-terminated surface of (Li$_{0.8}$Fe$_{0.2}$)OHFeSe. (**c-d**) Series of dI/dV spectra taken along the arrows shown in panel (**a-b**) respectively. The gray dashed lines indicate the position of the coherence peak at ±9 meV for (Li$_{0.8}$Fe$_{0.2}$)OHFeSe. (**e**) Topographic image of a single Zn adatom on a thick FeSe film. (**f**) Series of dI/dV spectra taken along the arrow shown in (**e**). The gray dashed lines indicate the position of the coherence peaks at ±2.5 meV for FeSe films.

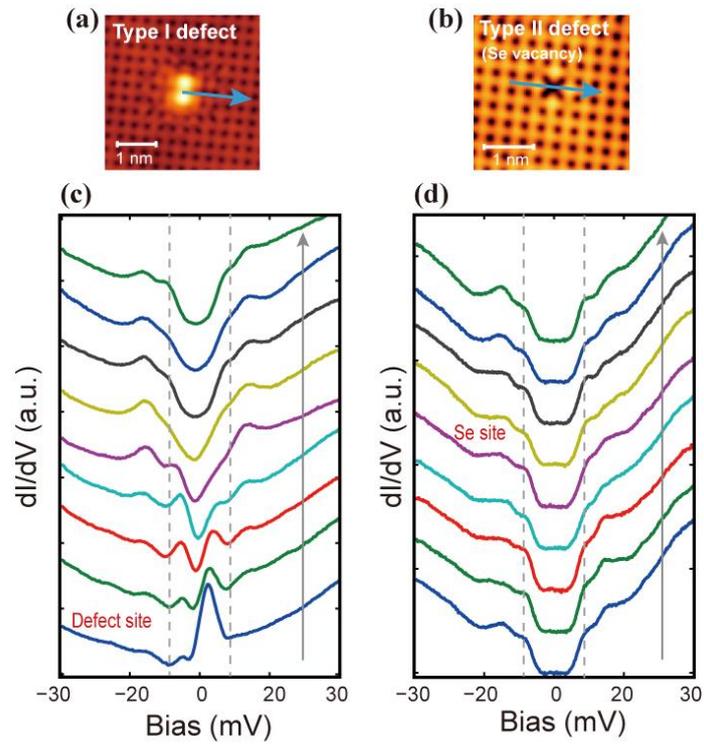

FIG. 9. Intrinsic-defect-induced effects on the superconductivity of $(Li_{0.8}Fe_{0.2})OHFeSe$. (**a-b**) Topographic images of type I defect (dimer-like defect on Fe site) and type II defect (Se vacancy) on an FeSe-terminated surface of $(Li_{0.8}Fe_{0.2})OHFeSe$. (**c-d**) Series of dI/dV spectra taken along the arrows shown in panel (**a-b**), respectively. The gray dashed lines indicate the position of the coherence peaks at ±9 meV.

# Supplementary Materials for

# Surface electronic structure and evidence of plain s-wave superconductivity in (Li$_{0.8}$Fe$_{0.2}$)OHFeSe

Y. J. Yan[1], W. H. Zhang[1], M. Q. Ren[1], X. Liu[1], X. F. Lu[3], N. Z. Wang[3], X. H.Niu[1], Q. Fan[1], J. Miao[1], R. Tao[1], B. P. Xie[1, 2], X. H. Chen[3], T. Zhang[1, 2*], D. L. Feng[1, 2*]

[1] State Key Laboratory of Surface Physics, Department of Physics, and Advanced Materials Laboratory, Fudan University, Shanghai 200433, China

[2] Collaborative Innovation Center of Advanced Microstructures, Fudan University, Shanghai 200433, China

[3] Hefei National Laboratory for Physical Science at Microscale and Department of Physics, University of Science and Technology of China, Hefei, Anhui 230026, People's Republic of China

## I. Vortex pinning on the FeSe-terminated surfaces and the "quantum limit"

We found that some of the vortices are pinned by type I defects, as demonstrated in Figs. s3a and s3b. In the ZBC mapping, black spots appear in the core center of pinned vortices (Fig. s3b), this is due to the core state being affected by the bound states induced by type I defects (see Fig. 9(c)). For this reason, we only studied vortices which were not pinned by defects (free vortices), as shown by white dashed circles in Fig. s3b. The data shown in Figs. 3(b-d) are taken on these vortices.

The "quantum limit" refers the situation where the thermal smearing is sufficiently low to realize discretized energy levels. For the vortex state in an *s*-wave superconductor, it requires $T/T_c < 1/(k_F\xi_0)$. Using $k_F = 0.21$ Å$^{-1}$ and $\xi_0 = 23$ Å obtained from our STM data and $T_c = 40$ K from transport measurements (Fig. s1), we get T < 8.2 K. The STM temperature was 4.2 K, which satisfies this criterion.

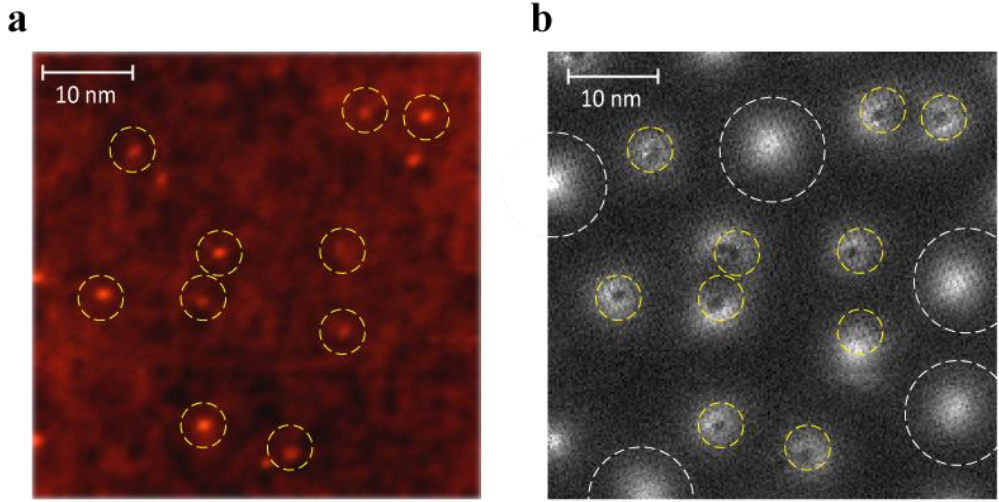

**Fig. s1 | (a)** STM topography of a 50×50 nm$^2$ area on the FeSe- terminated surface ($V_b$ = 50 mV, I= 50 pA), which is the mapping area of Fig. 3(a). The circled protrusions are type I defects. **(b)** ZBC mapping reproduced from Fig. 3(a), the yellow dashed circles mark vortices pinned by type I defects. The white dashed circles indicate free vortices which are not pinned by the defects.

## II. Quasi-particle interference (QPI) measurements

We preformed QPI measurements on both FeSe- and (Li$_{0.8}$Fe$_{0.2}$)OH- terminated surfaces, as displayed in Fig. s2 and Fig. s3. In Fig. s4, we compare the simulated FFT based on the unfolded BZ scenario and folded BZ scenario. The unfolded BZ scenario reproduces the QPI data better. Fig. s5 show the comparison of the dI/dV maps on FeSe- terminated surface and their FFTs taken at B = 0T and B = 11T.

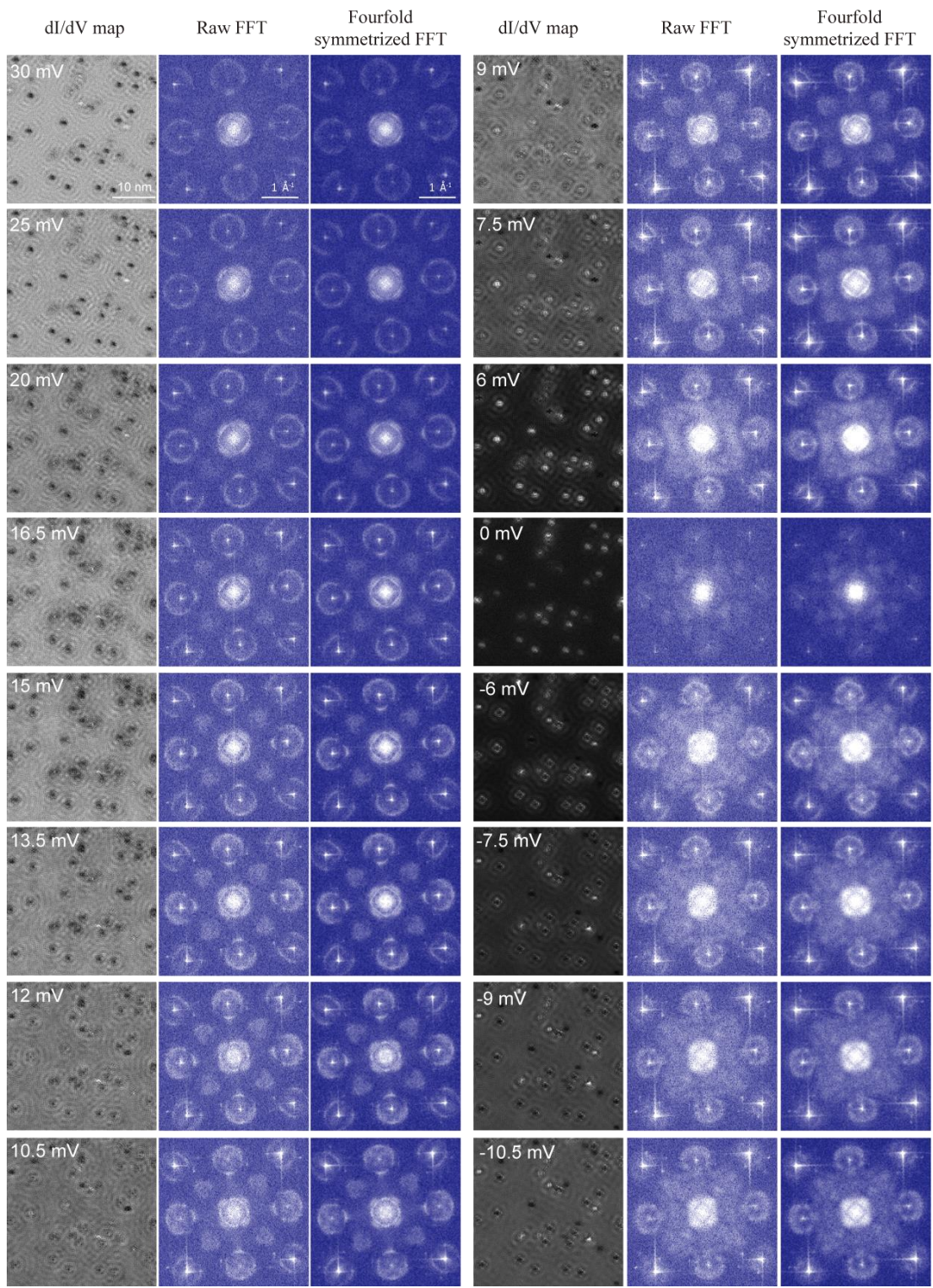

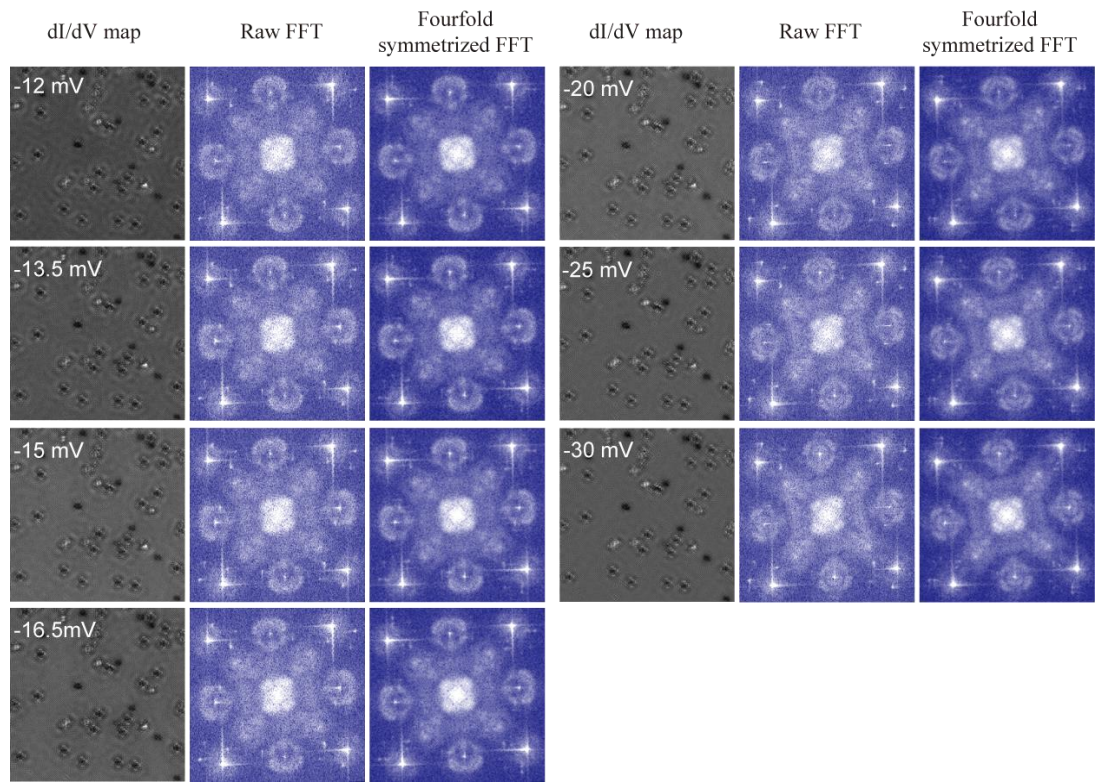

**Fig. s2** | dI/dV maps taken in a 35×35 nm² area of the FeSe-terminated surface, and their raw FFTs and four-fold symmetrized FFTs. All the dI/dV maps are taken at the set point of $V_b$ = 30 mV, I = 150 pA, $\Delta V$ = 1 mV. The mapping energies are labeled in the images. Each map has 300 ×300 pixels.

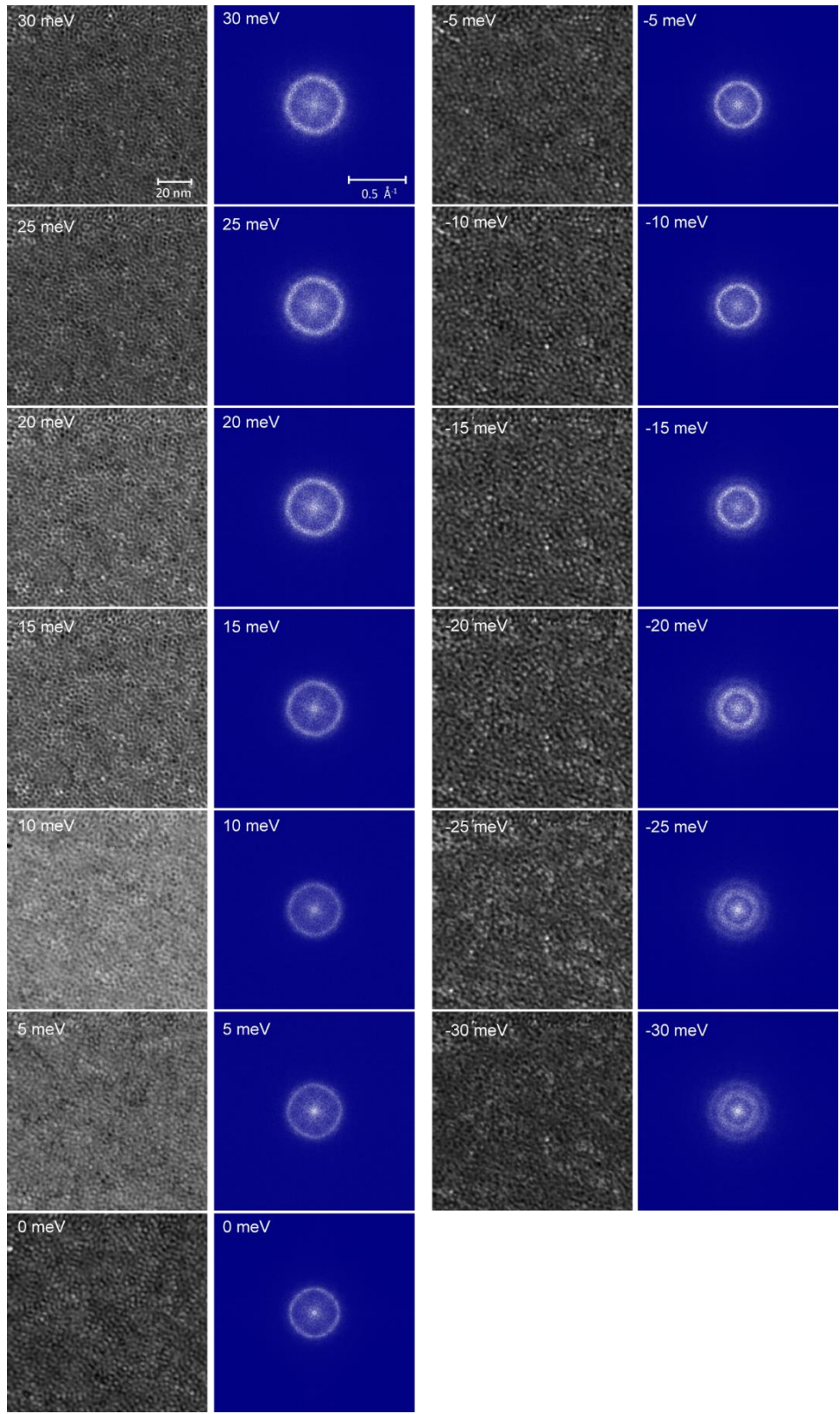

**Fig. s3** | dI/dV maps taken in a 120×120 nm² area of the $(Li_{0.8}Fe_{0.2})OH$-terminated surface and their FFTs (four-fold symmetrized). The mapping energies are labeled in the images. Each map has 400×400 pixels.

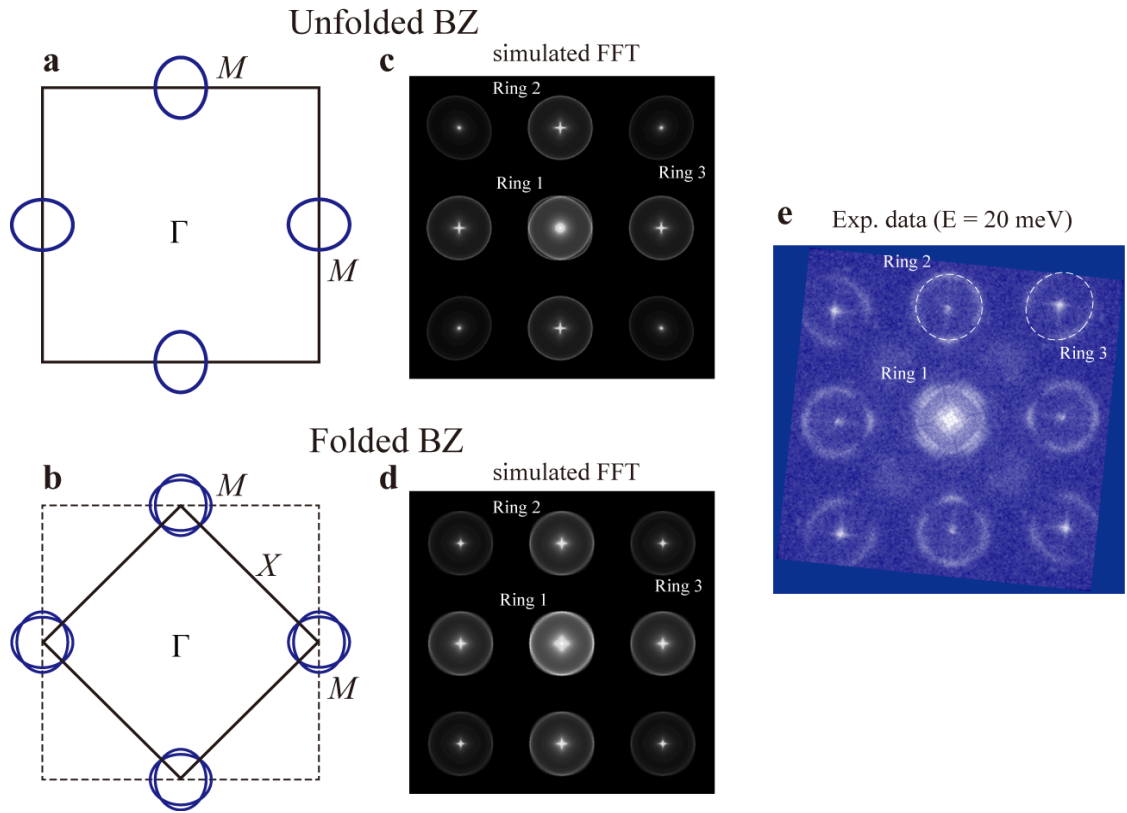

**Fig. s4 | Comparison of the unfolded BZ and folded BZ scenarios.** The Fermi surfaces of the FeSe layer are shown in **(a)** unfolded and **(b)** folded BZ scenarios. The electron pockets are shown with finite ellipticity. Note that in the folded BZ scenario all the different M points become equivalent. **(c)** and **(d)** Simulated FFT (JDOS) based on the unfolded BZ and folded BZ scenarios shown in panels **(a)** and **(b)**, respectively. **(e)** FFT image taken at E = 20 meV. One see that Ring 3 in panel **(e)** is elliptical, which is reproduced in panel **(c)** but not in panel **(d)**. Moreover, the four-fold anisotropy of Ring 1 is also better reproduced in panel **(c)**. These means the folding effect is likely weak in $(Li_{0.8}Fe_{0.2})OHFeSe$, at least for the FeSe- terminated surface.

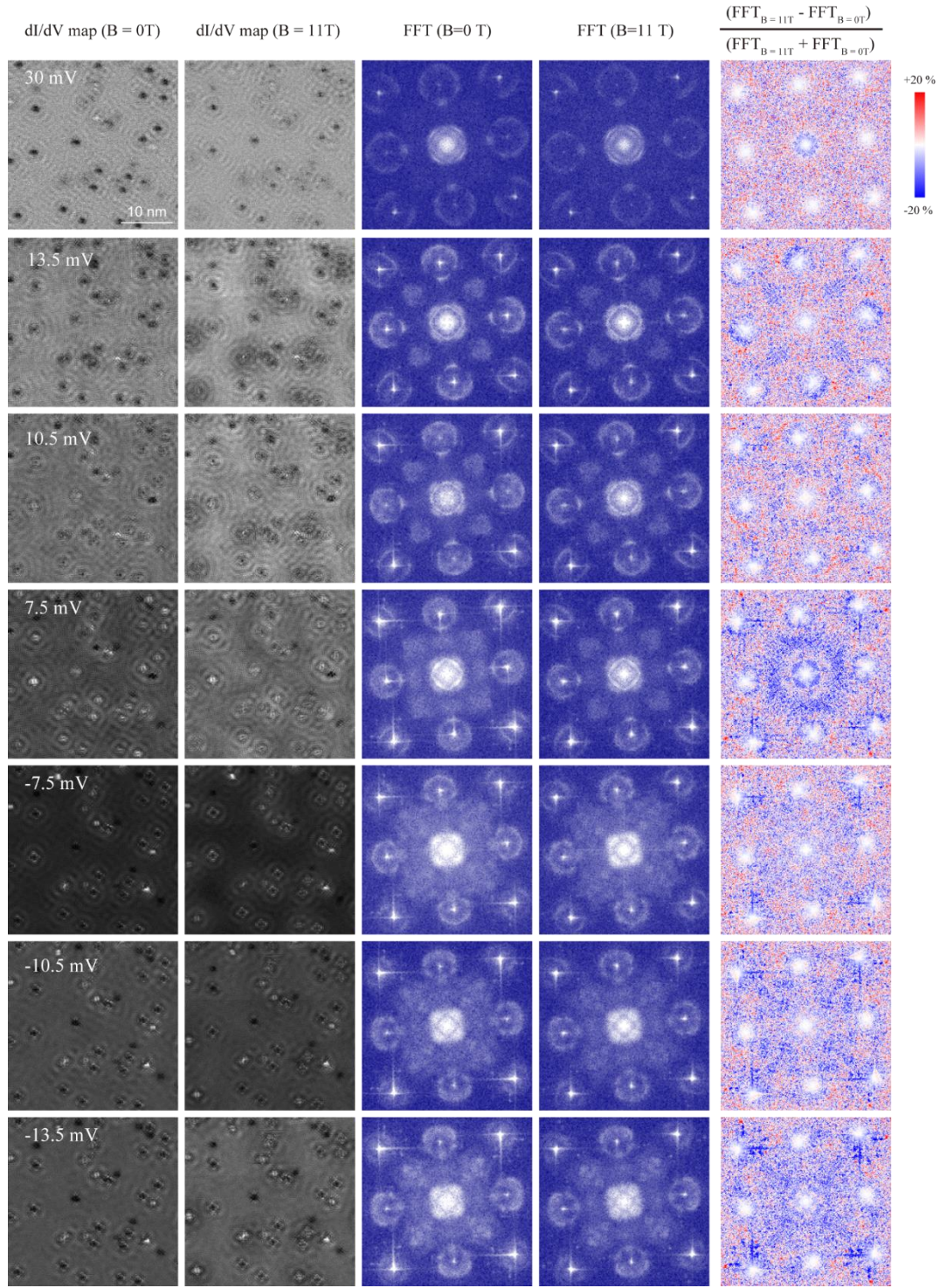

**Fig. s5 | Comparison of the dI/dV maps and their FFTs taken at B = 0T and B = 11T, at energies close to the superconducting gap edge.** An overall suppression of the scattering can been seen in the last column, which is calculated by $(FFT_{11T} - FFT_{0T}) / (FFT_{11T} + FFT_{0T})$. The intensity near (0,0) and all Bragg spots is suppressed by a factor of $1 - \Sigma[Gaussian(q_{(Bragg)}, \sigma)]$. All the dI/dV maps are taken at the same set point of $V_b$ = 30 mV, I = 150 pA, $\Delta V$ = 1 mV. The mapping biases are labeled in the first column.